# International Collaboration in Science and the Formation of a Core Group
*Journal of Informetrics* (forthcoming)

Loet Leydesdorff
University of Amsterdam, Amsterdam School of Communications Research (ASCoR)
Kloveniersburgwel 48, 1012 CX  Amsterdam, The Netherlands
loet@leydesdorff.net; http://www.leydesdorff.net

Caroline S. Wagner
SRI International
Arlington, Virginia, 22209, USA
Caroline.wagner@sri.com; http://www.cswagner.net

**Abstract**

International collaboration as measured by co-authorship relations on refereed papers grew linearly from 1990 to 2005 in terms of the number of papers, but exponentially in terms of the number of international addresses.  This confirms Persson *et al.*'s (2004) hypothesis of an inflation in international collaboration.  Patterns in international collaboration in science can be considered as network effects, since there is no political institution mediating relationships at that level except for the initiatives of the European Commission. Science at the international level shares features with other complex adaptive systems whose order arises from the interactions of hundreds of agents pursuing self-interested strategies.  During the period 2000-2005, the network of global collaborations appears to have reinforced the formation of a core group of fourteen most cooperative countries. This core group can be expected to use knowledge from the global network with great efficiency, since these countries have strong national systems.  Countries at the periphery may be disadvantaged by the increased strength of the core.

# 1. Introduction

An increasing share of scientific papers is co-authored by scientists from two or more nations. During the 1990s, a rapid rise occurred in internationally co-authored papers indicating a rise in collaboration (Doré *et al*., 1996; Georghiou, 1998; Glänzel, 2001). The increase was dramatic: international collaborations (as measured by internationally co-authored publications) doubled (Wagner & Leydesdorff, 2005a). The increase can be seen across all fields of science at more or less the same rate (Wagner, 2005). Collaboration continued to rise in the early 2000s. The number of internationally co-authored articles grew at a rate faster than traditional "nationally-co-authored" articles (NSB, 2002). Moreover, internationally co-authored articles appear to be cited more often than nationally co-authored papers (Narin, 1991; Persson *et al*., 2004).

We suggest that international collaboration in science can be considered as a communications network that is different from national systems and has its own internal dynamics (Gibbons *et al*., 1994; Price, 1963; Ziman, 1994; Katz & Hicks, 1998; Wagner, forthcoming; Wagner & Leydesdorff, 2005b). National systems have policies and institutions that mediate scientific communication, while at the global level the network exists primarily as a self-organizing system. The exception here is the European Union, where specific incentives exist to encourage formal international linkages among member countries. Does the EU emerge as an "international actor" or are different patterns (e.g., elite structures including the USA) more dominant?

## 2. Data and methodology

Data were harvested from the CD-Rom version of the *Science Citation Index* for articles, reviews, letters, and notes for 1990, 2000, and 2005.[1] In his study entitled *Evaluative Bibliometrics*, Narin (1976) proposed counting only articles, reviews, and notes as indicators of scientific performance. Braun *et al.* (1989) argued in favor of including letters as scientific output. However, the Institute for Scientific Information (ISI) no longer registered for the category of "notes" after 1995. We included 3,090 internationally coauthored notes in the counts for 1990 because the data was already organized in this format during a previous project. Table 1 provides an overview of the data for the three years in question.

| Year | (1) Nr of articles, reviews, letters, and notes | (2) Nr of addresses in documents in (1) | (3) Nr of authors in documents in (1) | (4) Internationally coauthored documents (a+r+l+n) | (5) Nr of addresses in (4) | (6) % Internationally coauthored documents |
|---|---|---|---|---|---|---|
| 1990 | 508,941 | 908,783 | 1,866,821 | 51,596 | 147,411 | 10.1 |
| 2000 | 623,111 | 1,432,401 | 3,060,436 | 121,432 | 398,503 | 19.4 |
| 2005 | 734,750 | 1,696,042 | 3,301,251 | 171,402 | 618,928 | 23.3[2] |

**Table 1**. Data on international collaboration comparing four years: 1990, 2000, and 2005.

For example, of the 1,011,363 records contained in the *Science Citation Index 2005*, only the 734,750 articles, reviews, and letters were considered. Among these documents, 171,402 were internationally coauthored; this is 23.3% of the total in column (1).

---

[1] Because we were surprised by our findings we repeated the complete analysis with 2006 data, but the results are not essentially different from those of 2005. The discussion thus focuses on these three years.
[2] The corresponding percentage for 2006 is 23.2%.

Collaboration was indicated by a co-authorship event at the document level. The country counts were done using integer counting, which attributes a count of "1" to each occurrence of authorship from a country among the set.[3] This leads to an asymmetrical matrix of documents versus countries. The cosines are computed on the basis of this matrix (Leydesdorff, 1989; Leydesdorff & Vaughan, 2006).

Since the distributions are not expected to be normal, it has been suggested that the cosine instead of the Pearson correlation is the proper measure for normalization (Ahlgren *et al.*, 2003; Boyack *et al.*, 2005; cf. Hamers *et al.*, 1989). The cosine normalizes to a geometric mean (rather than an arithmetic mean) and the consequent vector space model (Salton & McGill, 1983) is useful for the visualization of latent structures in the set. Since the cosine runs from zero to one, a very small number of relations can be expected to generate a cosine larger than zero. We considered cosine > 0.01 as a relevant threshold for discarding this incidental variation. Incidental variations may be caused by ongoing relations between supervisors and students or postdocs who have returned to their home countries.

The co-authorship events were additionally placed into a symmetrical matrix where country names appear on both axes, with the number of co-occurrence events appearing in the corresponding cell. Normalization, however, was based on the asymmetrical occurrence matrix because this matrix contains all information at the document level, including co-authorship relations among more than two countries. Leydesdorff & Vaughan (2006) showed that using the symmetrical co-occurrence matrix—which is based on multiplication of the asymmetrical occurrence matrix with its transposed—may lead to faulty conclusions about the underlying structure because information is lost. For example, correlations among co-occurrences can be spurious when based on multilateral co-occurrences at the document level (Waltman & Van Eck, 2007; Leydesdorff, 2007a).

---

[3] No effort was made to remove counts where the same author lists two different country addresses (Persson *et al.*, 2004). We assume that authors listing two addresses have colleagues in both institutions and therefore fall within a broad definition of collaboration.

Both the co-occurrence and the normalized tables were used to conduct network analysis using UCINET and Pajek software (De Nooy *et al.*, 2005). The normalized data can reveal structures such as resemblances in patterns which in non-normalized data are overshadowed by the effect of stars in the network with a high degree of centrality (e.g., the USA; Leydesdorff, 2007b). The results are presented below.

## 3. Results

Figure 1 visualizes the growth in international collaboration on the basis of the data provided in Table 1 above. The number of internationally coauthored publications has grown linearly ($r^2 > 0.99$). However, Figure 1 shows an exponential growth in the number of addresses of internationally collaborating authors ($r^2 > 0.99$), suggesting that the growth of the network extends to many more places around the globe, with a corresponding growth in the possibility of knowledge diffusion. The average number of addresses in any one internationally coauthored publication has grown from an average of 2.86 in 1990 to 3.61 in 2005, and this trend is accelerating.

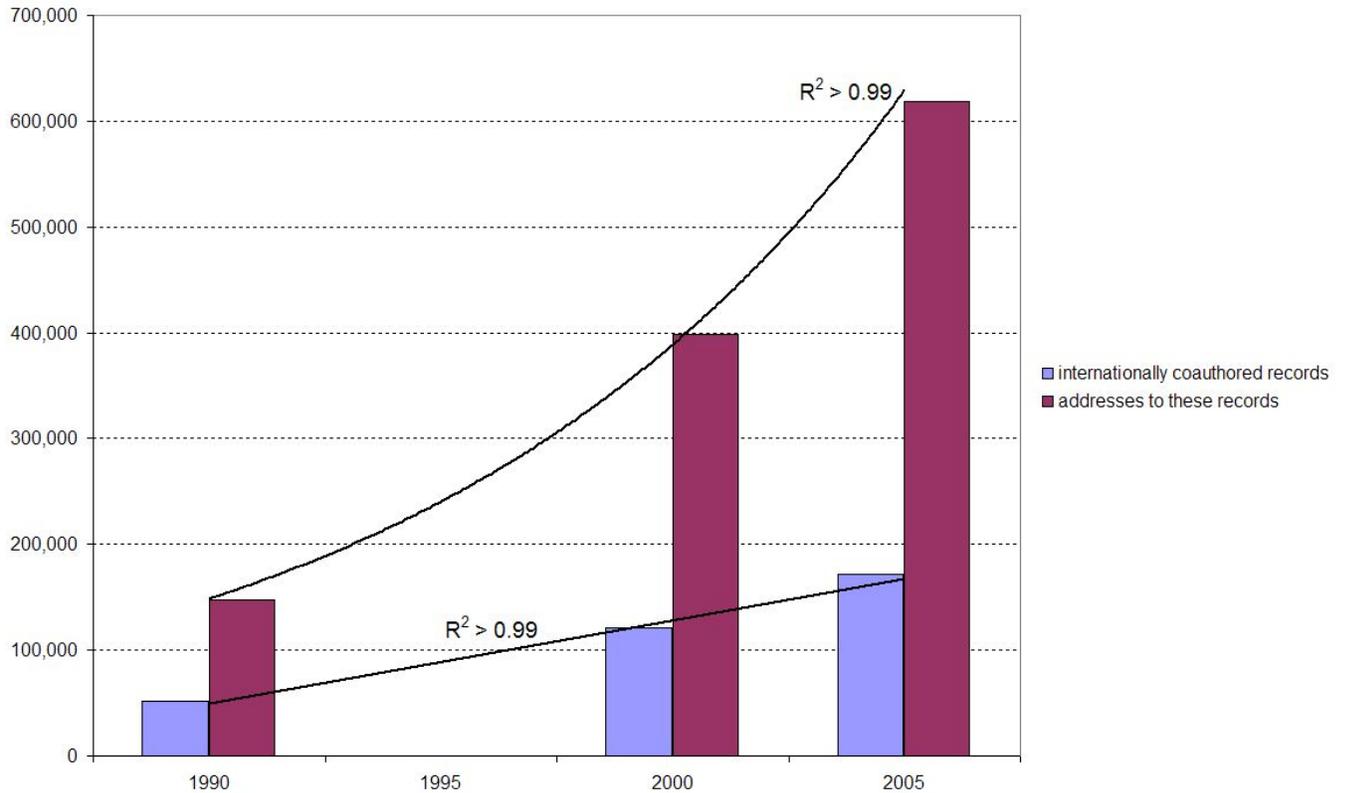

**Figure 1**. Trends in institutional participation in global science

In other words, internationally co-authored publications are increasingly multi-nationally co-authored. Because the number of records increases only linearly, Persson *et al*. (2004) concluded that this trend indicates an inflation in international collaborations.

Table 2 shows the results of the social network analysis for the global network of collaborations for the three years studied. Social network analysis provides us with a large number of statistics (Newman, 2000). First, the number of nodes in Table 2 represents the number of countries with authors participating in global science: this number increased by 20 between 1990 and 2000, with half of this growth due to the break-up of the Soviet Union into individual states, many of which began to participate in science as separate political entities during the decade of the 1990s. Between 2000 and 2005, the number of countries in the data set increased only from 192 to 194.

| Network statistics | *1990* | *2000* | *2005* |
|---|---|---|---|
| Number of nodes | 172 | 192 | 194 |
| Number of links | 1,926 | 3,537 | 9,400 |
| Size k-core component | 35 | 53 | 64 |
| Network density | 0.13 | 0.19 | 0.25 |
| Average degree | 22.4 | 36.9 | 48.6 |
| Average distance | 2 | 1.9 | 1.8 |
| Diameter | 3 | 3 | 3 |
| Graph betweenness | 0.26 | 0.16 | 0.14 |
| Average clustering coefficient | 0.78 | 0.79 | 0.79 |

**Table 2**. Social network analysis of the global science network

The number of links increased exponentially, as was already noted above. The network of international collaborations in science expanded more rapidly after 2000. The size of the *k*-core component [4]—that is, the dense network in the center—grew from 35 to 64 countries in this 15-year period.[5] The network as a whole became denser, which means that on average, countries participating in international collaboration are supporting an increasing number of collaborators at the global level. The average degree—a measure of the spread of influence across the network[6]—was higher in 2000 and 2005 than in 1990, suggesting that as the network grew, influence and power were spread more widely among nations at the global level (Burt, 2001).

The average distance across the network is the average number of steps it takes to go from a given node in the network to any other node in the network. Here the number in Table 2 shows that the number of steps between nodes is lower than two—a very low number in network terms—and that it decreased over the years examined. This suggests that the network is becoming more densely connected over time. The distance from one side of the network to the other is measured as the diameter, which is the number of steps at the global level, a low number that again confirms the density of the network. A small diameter in a large network is an indication of the number of connections that can be identified among the members. It also suggests the possibility of "small worlds" emerging within the network, due to the implicit connections possible between actors who are not in the same cluster.

---

[4] The core component in this case is the *k*-core, a subnetwork within a network where each node has at least *k* neighbors. *K* is a degree measure determined from an analysis based on the size of the entire network.
[5] The UK is counted as a single country in this table. The *Science Citation Index* provides address information for England, Scotland, Wales, and Northern Ireland, separately.
[6] The Freeman degree centralization measure expresses the degree of inequality or variance in a network as a percentage of that of a perfect star network of the same size (Hanneman & Riddle, 2006).

The final network measure shown in Table 2 is the average clustering coefficient. Clusters are groups within a network where redundant connections can be found. This coefficient measures the likelihood that nodes belong to a cluster. The global level can be considered as a single component, which means that all nations are connected to all other nations through some pathway in the network. Clustering occurs when there are many connections within a sub-set of a large network. For example, clustering is especially evident at the observed level within the European Union, where public policy has created incentives for international collaboration.

## 4. The Effects of Normalization

Normalization is needed to reveal the structure of the network. Because of its size alone, the USA dominates the network at the country level in every respect. Normalization enables us to consider the patterns of collaboration. When the data are normalized for the size of the participating countries using the cosine, the *k*-core measure reveals a latent structure in the data. It indicates that the central group in the network is becoming smaller (Figure 2).

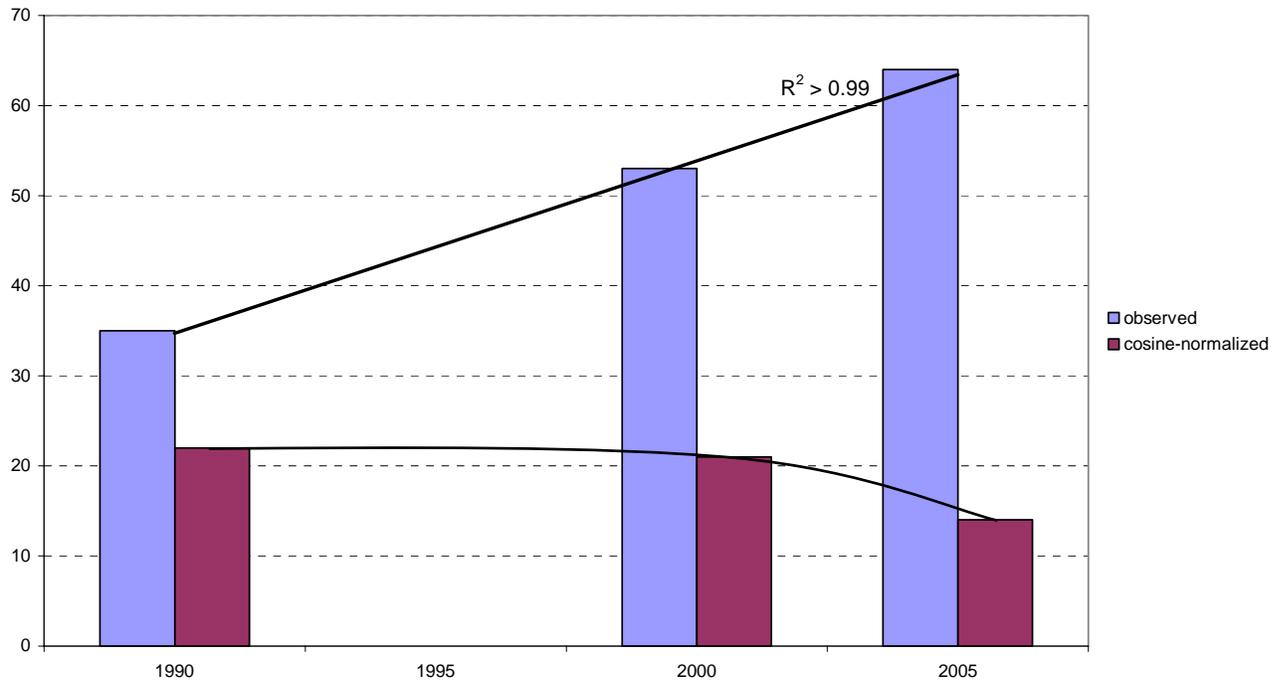

**Figure 2**. Normalized trends in participation in a core group of highly collaborative countries

In 2005, 14 advanced industrial nations belonged to the core group, down from 21 nations in 2000 and 22 in 1990. The other nations in the network are linked to this core group, but they are not bound to them in terms of *structural* relations. An analogy could be drawn to a volcano where the base is getting wider, but as it grows, it pushes the summit at the center higher and steeper.

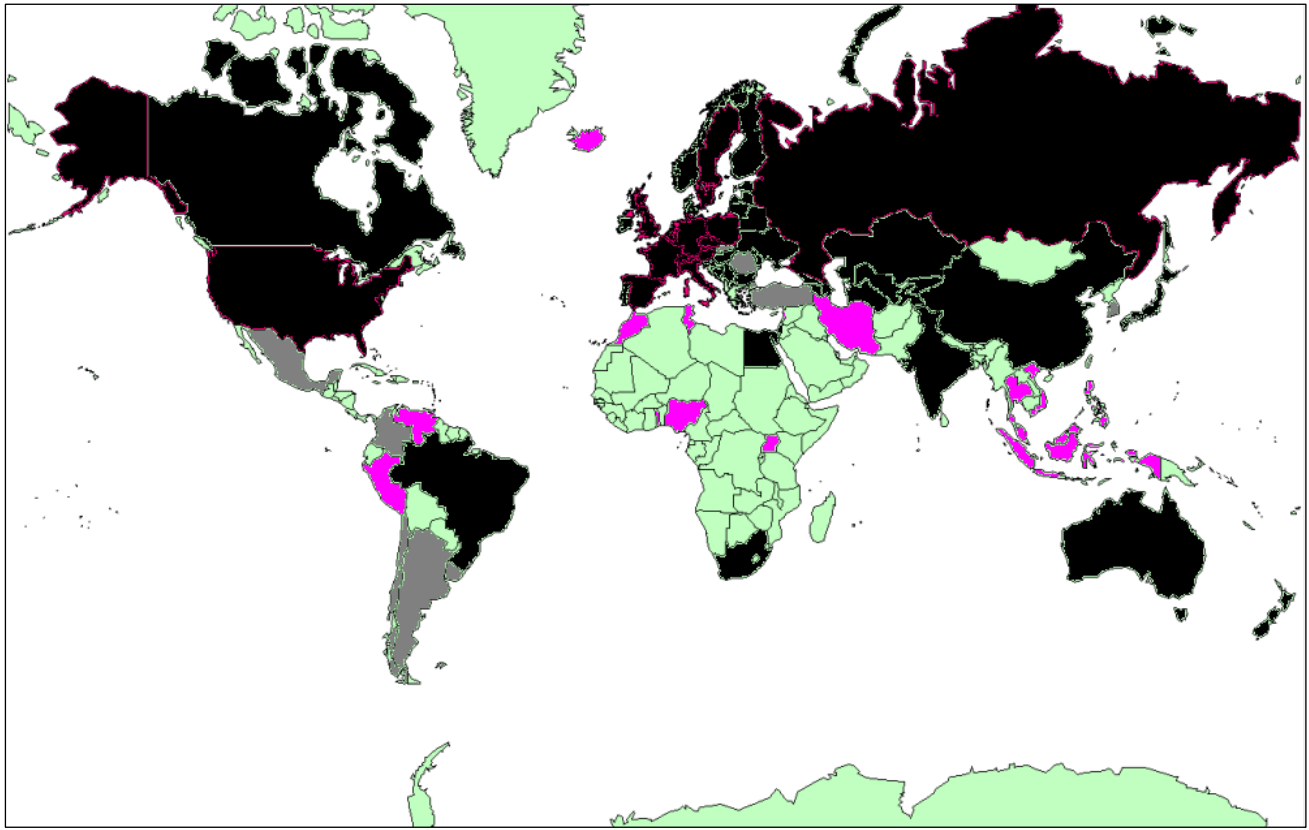

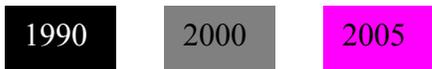

**Figure 3**. Observed participation in the *k*-core group of collaborating countries

Figures 3 and 4 compare membership in the *k*-core group of countries for the years studied before and after normalization. Countries shown in black were member of the core component of the observed network in 1990. The map shows the countries that joined this core group in 2000 and in 2005. The observed network is created by the occurrence of co-authors among scientists from different countries. This network has grown significantly and shows an increasing number of countries at the core.

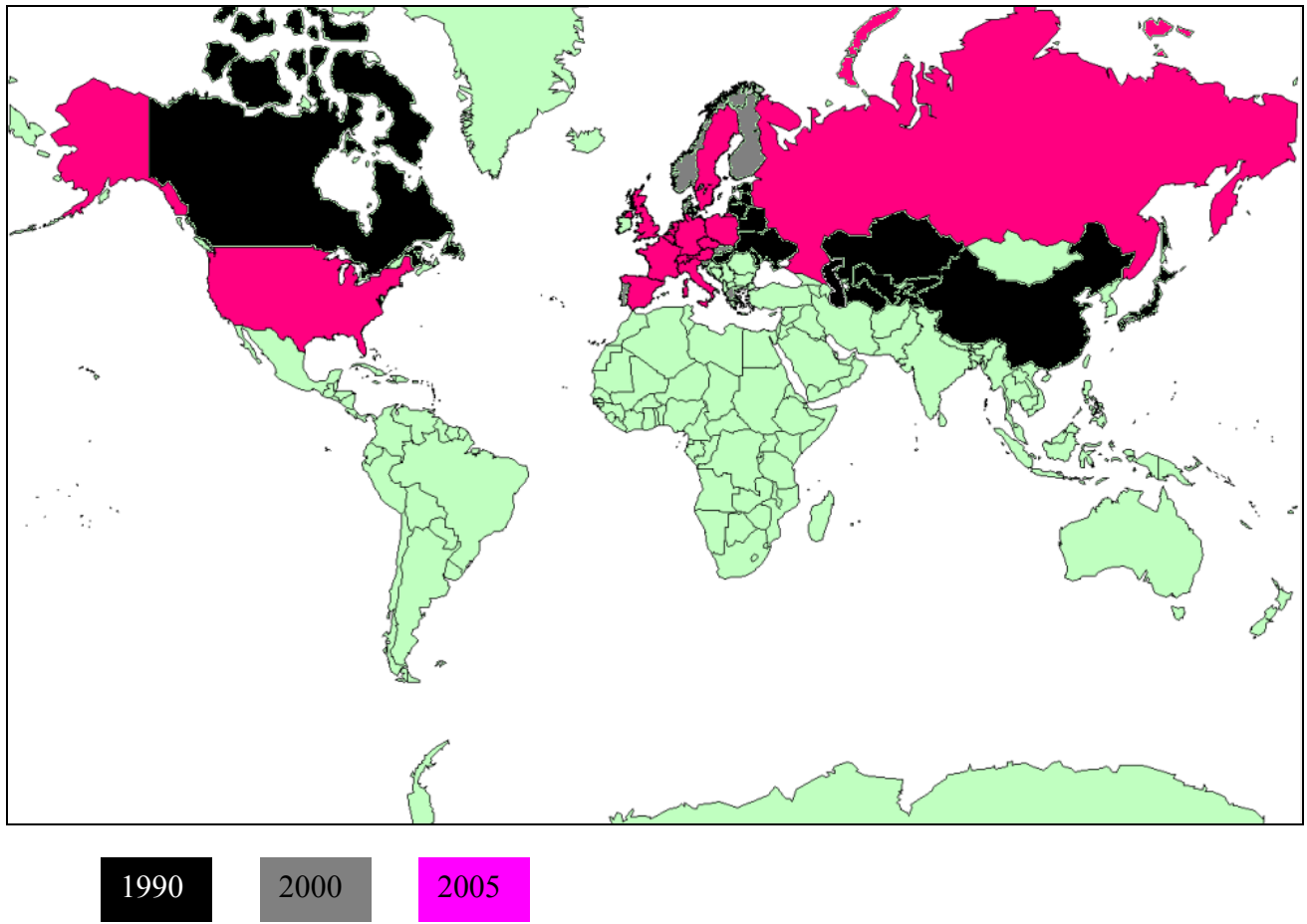

1990　2000　2005

**Figure 4**. Cosine normalized participation in the *k*-core group of collaborating countries

Figure 4 shows this network normalized for size. In this figure, the core group remains more or less stable from 1990 to 2000, but shrinks to a smaller set between 2000 and 2005. This suggests a network dynamic where a core group has evolved into a more tightly ordered and self-selective group. Note that by 2005 this core group no longer includes all EU-nations, but the remaining non-EU nations are only the USA, Russia, and Switzerland. This "Scienceland" does not coincide with "Euroland:" Denmark, Finland, and Portugal are no longer included, and Ireland has never been included (Leydesdorff, 2000).

We explored the network statistics for the normalized case as were done above (Table 2) for the non-normalized one. Interestingly, social network analysis provided us with the following indicator of change in the percentage of betweenness centrality among nations (Freeman, 1977, 1978; Leydesdorff, 2007):

| *1990* | | *2000* | | *2005* | |
|---|---|---|---|---|---|
| **USA** | 27.8 | *France* | 20.5 | *France* | 31.0 |
| *France* | 25.2 | **USA** | 15.3 | Russia | 9.6 |
| England [7] | 12.6 | England | 14.8 | UK | 7.8 |
| Australia | 5.0 | Russia | 6.1 | Sudan | 7.4 |
| South Africa | 4.4 | Germany | 5.3 | Colombia | 6.4 |
| Canada | 4.0 | Australia | 4.1 | Namibia | 6.2 |
| Egypt | 3.8 | Denmark | 3.9 | Germany | 6.2 |
| Fed Rep Ger | 3.4 | Canada | 3.5 | **USA** | 5.9 |
| Sweden | 3.1 | Malaysia | 3.3 | Senegal | 5.7 |
| Belgium | 2.9 | India | 3.2 | Cameroon | 5.0 |

**Table 3**: Rank order in the percentage of Betweenness Centrality, cosine ≥ 0.01

As noted, before normalization the USA was in every respect the most central country in all the years studied. After normalization, however, the USA lost its central position in terms of betweenness centrality during the 1990s to France. The demise of the Soviet Union made Russia an increasingly important player at the global level as well. Some of the developing countries have such a scattered pattern of international collaborations that they can demonstrate high betweenness centrality in one of the years for stochastic reasons. Among the major players, however, the tendencies visible in the 2000 data have been reinforced during the period 2000-2005 (Leydesdorff & Wagner, forthcoming).

---

[7] The data with addresses for England, Scotland, Wales, and Northern Ireland were not merged into a single field UK for 1990 and 2000 before the normalization (Wagner & Leydesdorff, 2005a).

## 5. Discussion

By adding a third year of data to our earlier analysis (Wagner & Leydesdorff, 2005a), we expected to find that international collaboration had continued to grow, that the network had become denser, and that more countries could be counted as part of the core component of the network. The first and second expectations are satisfied by the 2005 data: the network has grown larger and denser. The third expectation—that the core of the network structure has grown—is not supported after normalization. This finding suggests a different dynamic at work than we expected.

The core of any network holds a great deal of power in terms of how the periphery of a network organizes (Shils, 1998; Burt, 2001). In viewing the core of the network over the 15-year time period, a divergent outcome can be observed. At the observed level, the core of the network nearly doubles, suggesting a growing core group of collaborating countries. When the data are normalized, a smaller and tighter network is found at the core of the global system. In network terms, this suggests that the core group is becoming a more coherent cluster, perhaps reflecting more deliberate choices on the part of collaborators (and policymakers) to exploit the possibilities offered at the global level. In other words, as actors began to experience the phenomenon of globalizing links and distributed research during the 1990s, many of them shifted their choices to incorporate a wider view of the system. But those actors in the scientifically-advanced countries made more careful choices to limit their partners to specific countries.

During the 1990s, the "eco-system" was disturbed by changes in the political system such as the fall of the Soviet Union and the reunification of Germany. The introduction of the Internet led to a shift in communications from closed to open ("Mode 2") systems (Gibbons *et al.*, 1994). As these events changed the scientific system, the highly adapted entities (represented by the scientifically advanced countries) reorganized to take advantage of the changes and protect their positions. The core of the structural network then began to develop another order at the global level. Competition and cooperation shifted on the landscape, but this favored the highly adapted actors. The opportunities for knowledge diffusion are indeed greatly expanded at the global level, possibly benefiting scientists at the periphery in terms of having access to the core group. At the same time, the ability of the core group to access, absorb, and make use of participants from peripheral countries is made even greater.

In other words, the emergent pattern of the global system is not created as a result of the actions or plans of a single entity or actor in the system. The order arises from rules embedded at the level of the researchers themselves, and self-organizes through collective action. In evolutionary dynamics, the early phase of the shift that occurred in the 1990s generated variation, and the actors within the system responded to stabilize the changes, as retention mechanisms were put into place. International collaborations have become part of the system that now includes local, regional, national, and global levels of order.

## 6. Implications for Research Policy and Management

Public funding for science has been supported and managed in the interests of nations for more than a century (Price, 1963; Ziman 1994). Government investment in and creation of important technologies spurred economic growth and reinforced a model for postwar science policy at the national level in the United States and later in Europe and Japan. Large federal agencies grew up to manage the relationships between the political and scientific communities.[8]

---

[8] In the public realm, these kinds of institutions included those dedicated to provide basic science funding. The U.S. National Science Foundation is one example. Other agencies were mission-oriented, but had a significant basic and applied research and development budget. The National Institute of Health or the Defense and Energy Departments in the U.S. are examples of these agencies.

When the scale or scope of research stretched the budgets of the scientifically advanced nations, joint investments in megascience projects such as the International Space Station or the Large Hydron Collider resulted in multi-national projects, but political accountability for investment remained at the national level (Galison & Hevly, 1992).

The growth of the global network of science does not mean we are witnessing the death of the nation-state or even a reduction in its influence in scientific investments. However, it does mean that nations must take careful stock of the conduct of science at the global level as well as at the national and regional levels. As the system expands, useful innovation can increasingly occur somewhere else; identifying innovations and making them locally available will be a major challenge for policymakers. Finding ways to evaluate distributed scientific research and local absorptive capacity is another.

The global system requires new approaches to public accountability and evaluation. The system as a whole is likely gaining in efficiency by distributing tasks and sharing resources, as opposed to creating redundant capacities in different countries. Distributed tasking of scientific research within highly complementary, competitive, and self-directed teams can accelerate the testing of ideas and the validation of scientific concepts. Nevertheless, it may become increasingly difficult to track spending to outputs and outcomes, which has been the model for much of public accountability for science in the past.

As scientific capacity continues to grow around the world, and more links are made among countries, the flow of knowledge among them may also grow. Nevertheless, if a core group is indeed developing its own identity and enhancing its absorptive capacity within a global system, developing countries may find that good ideas will flow from their laboratories to the larger actors who are better able to publish these ideas in scientific journals. It may be that the benefits of science can be disseminated more effectively to the periphery, but this may require deliberate policy actions on the part of the countries in the center. Managing a complex, open system requires crafting new incentives to encourage knowledge flows and participation that favors the peripheral members.

ACKNOWLEDGEMENT

Caroline Wagner gratefully acknowledges support from the U.S. Department of Energy Office of Basic Science. The authors wish to thank Susan A. Mohrman for comments on a previous draft.